\documentclass[a4paper]{article}
\usepackage[affil-it]{authblk}

\usepackage{graphicx}
\usepackage{graphics}
\usepackage{amsmath,amssymb}
\usepackage[usenames]{color}
\usepackage{subfigure}
\usepackage{hyperref}

\newcommand{\be}{\begin{equation}}
\newcommand{\ee}{\end{equation}}
\newcommand{\bea}{\begin{eqnarray}}
\newcommand{\eea}{\end{eqnarray}}
\newcommand{\ben}{\begin{enumerate}}
\newcommand{\een}{\end{enumerate}}
\newcommand{\bit}{\begin{itemize}}
\newcommand{\eit}{\end{itemize}}

\definecolor{BrickRed}{cmyk}{0,0.89,0.94,0.28}
\definecolor{MidnightBlue}{cmyk}{0.98,0.13,0,0.43}
\definecolor{DarkGreen}{rgb}{0.100806,0.495968,0.209979}
\definecolor{orange}{rgb}{0.587167,0.354498,0.146197}

\begin{document}

\title{Comment on  `` Second Law of Thermodynamics without Einstein Relation''}

\author[]{Robert Alicki\thanks{E-mail: \texttt{robert.alicki@ug.edu.pl}}}

\affil[]{ICTQT, University of Gda\'nsk, 80-952 Gda\'nsk, Poland}

\date{\vspace{-5ex}}

\maketitle


\begin{abstract}
It is argued the the idea of a single temperature-like variable, introduced in \cite{Sorkin}, which enters a generalized  second law for Markovian open system in non-equilibrium environment is not sufficient for a consistent and useful thermodynamic formalism. The origin of the Second Law and ``local temperatures''  in Markovian theory of Quantum Open Systems is revisited and illustrated by simple examples.
\end{abstract}


The authors of \cite{Sorkin} used, for their construction, classical overdamped Langevin equations and stochastic thermodynamics, while here, the formalism of Quantum Thermodynamics and Quantum Markovian Master Equations (see e.g. \cite{QT-review} for the references) is applied. It is much simpler (also notationally), easier to derive from the first principles and contains as limit cases classical results.
\par
Consider the following Markovian Master Equation for the reduced density matrix $\rho(t)$ of a quantum open system in a general non-equilibrium time-dependent environment   ($\hbar = k_B =1$)
\begin{equation}
\frac{d}{dt} \rho (t) =\mathcal{L}(t) \rho (t), \quad t\geq 0.
\label{MME}
\end{equation}
Markovian assumption means, by the definition,  that  $\mathcal{L}(t)$ is Gorini -Kossakowski -Lindblad - Sudarshan (GKLS) generator \cite{nonM}. The monotonicity of the relative entropy $S(\rho |\sigma) \equiv \mathrm{Tr}\bigl(\rho[\ln \rho - \ln \sigma)]\bigr)$  with respect to any  Markovian evolution implies the following entropy balance 
\begin{equation}
\frac{d}{dt}S(t) - \sum_{\alpha}{J^S}_{\alpha}(t) \geq 0 , 
\label{SIIlaw}
\end{equation}
where $S(t) = \mathrm{Tr}\bigl(\rho(t) \ln \rho(t) \bigr)$ is von Neumann entropy identified with the thermodynamical entropy and the entropy currents
\begin{equation}
{J^S}_{\alpha} (t)= -\mathrm{Tr}\Bigl[\bigl(\mathcal{L}_{\alpha}(t)\rho (t) \bigr)\ln \sigma_{\alpha}(t)\Bigr], 
\label{Scurrent}
\end{equation}
are defined by the decomposition  of  $\mathcal{L}(t) $ into a sum of  GKLS generators with different temporal stationary states  $\sigma_{\alpha}(t)$
\begin{equation}
\mathcal{L}(t) =\sum_{\alpha}\mathcal{L}_{\alpha}(t) , \quad \mathcal{L}_{\alpha}(t) \sigma_{\alpha}(t) = 0 .
\label{decomposition}
\end{equation}
Although \eqref{SIIlaw} looks like a plausible formulation of the Second Law for open systems it  is still ambiguous and needs further physical inputs. The decomposition of  $\mathcal{L}(t)$ is generally nonunique and the definition of entropy currents must be connected to the definition of heat currents ${J^H}_{\alpha} (t)$  and  temperatures, ${J^S}_{\alpha} (t) = {J^H}_{\alpha} (t) / T_{\alpha}(t)$, satisfying a proper energy balance (First Law)
\begin{equation}
\frac{d}{dt}E(t) = \sum_{\alpha}{J^H}_{\alpha}(t)  -  P(t), 
\label{Ilaw}
\end{equation}
with the internal energy $E(t)$ and the produced power $P(t)$. 
\par
The authors of \cite{Sorkin} define a single temperature characterizing Markovian open system with nonequilibrium environment
\begin{equation}
T(t) = \frac{J^H(t) }{J^S(t) } 
\label{temp}
\end{equation}
using a classical version of  \eqref{MME}, \eqref{SIIlaw} with a trivial decomposition \eqref{decomposition}. This cannot be a physically meaningful formulation as only the case of a single equilibrium bath can be characterized by a single temperature. 
\par
To illustrate the need for various types of ``local temperatures'' in nonequilibrium situations we begin with the case of quantum open system with a slowly varying Hamiltonian $H(t)$ weakly coupled to several heat baths at temperatures $T_{\alpha}$, each yielding its own $\mathcal{L}_{\alpha}$. Then one can show \cite{QT-review} that \eqref{SIIlaw} and \eqref{Ilaw} are satisfied with  $E(t) = \mathrm{Tr}\bigl(\rho (t) H(t)\bigr)$,  $P(t) = -\mathrm{Tr}\bigl(\rho (t) \frac{d}{dt}H(t)\bigr)$ and ${J^H}_{\alpha} (t)= \mathrm{Tr}\bigl(H(t)\mathcal{L}_{\alpha}\rho(t)\bigr)$.

The second example is a stationary but nonequilibrium photon bath at Gaussan state fully characterized by the frequency dependent population $n(\omega)$. One can introduce the ``local temperature'' $T_{\omega}$ defined by
\begin{equation}
\frac{n(\omega)}{n(\omega) + 1} = e^{-\omega/T_{\omega}}.
\label{loctemp}
\end{equation}
Dynamics of the ``atom'' weakly interacting with such a bath can be described by the Markovian Master Equation with the generator decomposed into GKLS generators labelled by the relevant  Bohr frequencies $\mathcal{L}=\sum_{\omega}\mathcal{L}_{\omega}$
\begin{equation}
\mathcal{L}_{\omega} = \gamma_{\omega}\bigl(\mathcal{D}[S_{\omega}] + e^{-\omega/T_{\omega}} \mathcal{D}[S^{\dagger}_{\omega}] \bigr). 
\label{decomposition1}
\end{equation}
Here , $\gamma_{\omega} >0$ and the short-hand notation for GKLS generators means $\mathcal{D}[A] \rho = A \rho A^{\dagger} - 1/2\bigl(  A^{\dagger} A \rho + \rho A^{\dagger} A$. The operators $S_{\omega}$ ($S^{\dagger}_{\omega}$) describe downward (upward) transitions between energy levels separated by $\omega$. Therefore, the stationary state for \eqref{decomposition1} is a Gibbs state $\sigma_{\omega} = Z^{-1} e^{-H/T_{\omega}}$ at the local temperature $T_{\omega}$. 
\par
Similar structure appears for periodically driven (with frequency $\Omega$)  open systems. 
Bohr frequencies are replaced by a set $\{\bar{\omega} + q \Omega , q = 0, \pm1, \pm 2,...\}$, where $\bar{\omega}$ are Bohr frequencies for the ``averaged Floquet Hamiltonian'' \cite{localT}. 
\par
For the examples discussed above the relations  \eqref{SIIlaw}  - \eqref{Ilaw} provide a consistent thermodynamic theory applicable to non-equilibrium systems  and allowing, for example, to derive Carnot-like bounds on engine efficiences. Natural applications of this formalism are photovoltaic, thermoelectric, electrochemical cells and their biological counterparts (see \cite{AlGKLS} and references therein).

\end{document}